\documentclass[12pt]{article}
\usepackage{amssymb}
\usepackage{epsfig}
\begin{document}

\title{
Spectral Properties of the $k$--Body Embedded Gaussian
Ensembles of Random Matrices for Bosons\\}

\author{ T. Asaga\footnote{Present address: Laboratory of Physics,
College of Science and Technology, Nihon University, 7-24-1
Narashino-dai, Funabashi, Chiba, Japan 274-8501}, 
L. Benet, T. Rupp and H. A. Weidenm\"uller\\ 
Max--Planck--Institut f\"ur Kernphysik \\
D--69029 Heidelberg, Germany\\}

\date{\today}

\maketitle

\begin{abstract}
  We consider $m$ spinless Bosons distributed over $l$ degenerate
  single--particle states and interacting through a $k$--body random
  interaction with Gaussian probability distribution (the Bosonic
  embedded $k$--body ensembles). We address the cases of orthogonal
  and unitary symmetry in the limit of infinite matrix dimension,
  attained either as $l \to \infty$ or as $m \to \infty$. We derive an
  eigenvalue expansion for the second moment of the many--body matrix
  elements of these ensembles. Using properties of this expansion, the
  supersymmetry technique, and the binary correlation method, we show
  that in the limit $l \to \infty$ the ensembles have nearly the same
  spectral properties as the corresponding Fermionic embedded
  ensembles. Novel features specific for Bosons arise in the dense
  limit defined as $m \to \infty$ with both $k$ and $l$ fixed. Here we
  show that the ensemble is not ergodic, and that the spectral
  fluctuations are not of Wigner--Dyson type. We present numerical
  results for the dense limit using both ensemble unfolding and
  spectral unfolding. These differ strongly, demonstrating the lack of
  ergodicity of the ensemble.  Spectral unfolding shows a strong
  tendency towards picket--fence type spectra. Certain eigenfunctions
  of individual realizations of the ensemble display Fock--space
  localization.\\

\end{abstract}

\newpage

\section{Introduction}
\label{int}

Random--matrix theory (RMT) was introduced by Wigner~\cite{wig56} to
describe the complexity of the spectra and wave functions of atomic
nuclei. It was soon realized that RMT could be applied successfully to
a large variety of systems~\cite{bro81,guh98}. Examples include atoms,
molecules, atomic nuclei and quantum dots. For such systems, the RMT
modeling is not completely realistic since all many--body systems are
effectively governed by one-- and two--body forces. This fact
motivated work on the two--body random ensembles for Fermions~(see
Refs.~\cite{fre71} to~\cite{bro76}) and inspired the pioneering work
of French and collaborators on the subject. Especially, it led to the
introduction of the $k$--body embedded ensembles by Mon and
French~\cite{mon75} (see also Ref.~\cite{ver84}). In the embedded
ensembles (EE), stochasticity is generated at the $k$--body level. The
matrix of the $k$--body interaction is then taken between
$m$--particle states with $k \le m$.  The latter are obtained by
distributing $m$ particles over $l$ degenerate single--particle
levels. For $k < m$, the matrix elements of the $k$--body interaction
in the space of $m$--body states are correlated. The number of
independent random variables is much smaller than in RMT. The question
arose whether these more realistic ensembles yield the same results as
RMT. Early numerical experiments performed for interacting Fermion
systems~\cite{fre71,boh71,boh71a} of rather small matrix dimension
showed that spectral fluctuation properties of the EE agreed with
those of RMT. For Bosonic systems, Manfredi~\cite{man84} argued that
the spectral fluctuations should coincide with the predictions of
RMT. More recently Patel {\it et al.}~\cite{patel00} performed
simulations for a Bosonic case with $m > l$ and concluded that the
agreement with RMT carries over into the so--called dense limit. As
far as we know, there are no results on the ergodic behavior of the EE
for Boson systems.

Interest in model Hamiltonians with random two--body interactions has
resurged in recent years in several areas of many--body physics (see
Refs.~\cite{flam96} to~\cite{kot01}), and the question of possible
further differences between such models and RMT has resurfaced. For
instance, Altshuler {\it et al.}~\cite{alt97} suggested localization
in Fock space as a realistic possibility in a system of interacting
electrons in a random one--body potential. In a different context,
calculations using random two--body interactions within the nuclear
shell--model~\cite{joh98} and in the framework of the Interacting
Boson Model~\cite{bij99} led to the observation of pairing properties
and band structures in the spectra. These ``regular'' features in the
low--lying part of the spectrum are robust and seem to contradict the
expectations based on the canonical ensembles of RMT. Recently, three
of the present authors found a novel analytical approach to the
Fermionic embedded ensembles~\cite{ben01a,ben01b} in the limit of
infinite matrix dimension ($l\to \infty$). The main results of this
approach are: (i)~Within a certain parameter range ($2k > m$), the
average spectrum has semicircle shape, and the spectral fluctuation
properties coincide with those of RMT. (ii)~The spectral density
changes shape at or near $2k = m$ and becomes Gaussian. (iii)~In the
dilute limit ($k \ll m \ll l$) the spectral fluctuations are
completely uncorrelated (Poissonian). (iv)~The spectral fluctuations
change gradually from Wigner--Dyson to Poissonian.

In this paper, we extend our analytical treatment to the case of the
$k$--body embedded Gaussian ensembles of random matrices for Bosonic
systems. Our motivation for doing so is twofold. First, it is
instructive to understand possible differences arising from symmetry
(fully symmetric versus fully antisymmetric states). To this
theoretical argument we add a second and more practical consideration:
In an atomic trap, Bosonic atoms occupy partly degenerate
single--particle states. The interaction will lift the degeneracy. A
random--matrix approach should reveal the generic features of the
resulting spectrum.

Our approach is largely patterned after the one for Fermionic systems
which in turn is based on the invariance properties of the EE. We
focus attention on the second moment $B^{(k)}$ of the random $k$--body
interaction in the Hilbert space of many--body states. We derive an
eigenvector expansion formally equivalent to the Fermionic one. We
exploit this formal equivalence and obtain results for the shape of
the spectral density and for the spectral fluctuations in terms of the
parameters $k$, $m$, and $l$ of the problem. All our results are obtained
in the limit of infinite Hilbert--space dimension. For Bosons this
limit is realized either by letting $l \to \infty$ (the same limit as
for Fermions), or by letting $m \to \infty$ with $k$ and $l$ fixed.
This second case is novel and has no analogue in the Fermionic case.
We refer to this case as to the dense limit. The tools at our disposal
are the supersymmetry method~\cite{efe83,ver85} and the binary
correlation method. Neither of these methods really works in the dense
limit except to show that standard Wigner--Dyson statistics will not
apply. Therefore, we resort here to numerical simulations. 

The Bosonic embedded ensembles are defined in such a way that for $k =
m$, they coincide with one of the standard ensembles of RMT.
Deviations can occur only for $k < m$. In the dilute limit ($k \ll m
\ll l$) there is no difference between Bosons and Fermions. Therefore,
we expect that the embedded ensembles for Fermions and Bosons are
qualitatively similar as long as $k < m < l$. This is indeed what we
find. The physically most interesting and novel case is, therefore,
the dense limit. Here, we prove analytically that the ensemble is not
ergodic. Numerical results for the spectral correlations are obtained
both by ensemble unfolding and by spectral unfolding. In the latter
case we find a strong tendency of the spectra towards picket--fence
behavior. We also show that certain eigenfunctions display Fock--space
localization.

The paper is organized as follows. The Bosonic embedded ensemble are
defined in Section~\ref{def}. The second moment $B^{(k)}$ is
introduced, and an important identity is derived, in Section~\ref{bas}.
The eigenvector expansion derived in Section~\ref{eig} is a central
analytical result of this work. This expansion is formally identical
to the one found for the Fermionic ensembles. It enables us to
calculate the low moments of the interaction (Section~\ref{low}) and
to apply the supersymmetry method (Section~\ref{sup}) in a
straightforward way. In Section~\ref{num} we present numerical results.
Section~\ref{con} contains our conclusions. A preliminary account of
some of our results was given in Ref.~\cite{asa01}.

\section{Definitions and Elementary Facts}
\label{def}

We consider $m$ spinless Bosons in $l$ degenerate single--particle
states with associated creation and annihilation operators
$b_j^{\dagger}$ and $b_j$ where $j = 1,\ldots,l$. The Bosons are
coupled through a random $k$--body interaction $V_k(\beta)$, with $k
\leq m$, given by
\begin{equation}
\label{eq2.1}
V_k(\beta) = \sum_{{1 \leq j_1 \leq j_2 \leq \ldots \leq j_k \leq l}
  \atop {1 \leq i_1 \leq i_2 \leq \ldots \leq i_k \leq l}}
  v_{j_1,\ldots,j_k; i_1,\ldots,i_k} \,
  {b_{j_1}^{\dagger} \ldots b_{j_k}^{\dagger} b_{i_k} \ldots b_{i_1}
  \over {\cal N}(j_1,\ldots,j_k) {\cal N}(i_1,\ldots,i_k) } \ .
\end{equation}
We refer to $k$ as to the rank of the interaction. As in the canonical
case, we use the labels $\beta = 1$ and $\beta = 2$ for the orthogonal
and the unitary ensemble, respectively. The matrix element
$v_{j_1,\ldots,j_k; i_1,\ldots,i_k}$ of the $k$--body interaction
taken between the single--particle states $j_1,\ldots, j_k$ and
$i_1,\ldots, i_k$ is totally symmetric with respect to $j_1,\ldots,
j_k$ and $i_1,\ldots, i_k$. The elements differing in the sequence of
indices $\{j_1 \ldots j_k; i_1 \ldots i_k \}$ (except for permutations
of $\{j_1 \ldots j_k\}$ and of $\{i_1 \ldots i_k \}$ and for
symmetries specified by $\beta$) are uncorrelated
Gaussian--distributed random variables with zero mean and a common
second moment $v_0^2$. Thus,
\begin{eqnarray}
\label{eq2.2}
\overline{v_{j^{}_1,\ldots,j^{}_k; i^{}_1,\ldots,i^{}_k}
    v_{j'_1,\ldots,j'_k; i'_1,\ldots,i'_k}} &=& v_0^2 [\delta_{j^{}_1
    i'_1} \ldots \delta_{j^{}_k i'_k} \delta_{i^{}_1 j'_1} \ldots
    \delta_{i^{}_k j'_k} \nonumber \\
&&\qquad + \delta_{\beta 1} \delta_{j^{}_1 j'_1} \ldots
    \delta_{j^{}_k j'_k} \delta_{i^{}_1 i'_1} \ldots \delta_{i^{}_k
    i'_k}] \ .
\end{eqnarray}
The overbar denotes the average over the ensemble. Without loss of
generality we put $v_0^2=1$ in the sequel. The normalization
coefficients ${\cal N}(j_1,\ldots,j_k)$ and ${\cal N}(i_1,\ldots,i_k)$
in Eq.~(\ref{eq2.1}) are defined below and are introduced in order to
obtain for $k=m$ the canonical ensembles of random--matrix theory.
The number of independent random variables is given by $K_{\beta} =
\beta {l+k-1 \choose k}[{l+k-1 \choose k }+\delta_{\beta 1}]/2$. This
defines the Bosonic $k$--body embedded Gaussian orthogonal (unitary)
ensemble of random matrices, respectively, in short BEGOE($k$) and
BEGUE($k$).

Hilbert space is spanned by the orthonormal $m$--particle states $|
\mu \rangle$ given by
\begin{equation}
\label{eq2.3}
| \mu \rangle = { b_{j_1}^{\dagger} \ldots b_{j_m}^{\dagger} \over
  {\cal N}(j_1,\ldots,j_m) } \, | 0 \rangle \ ,
\end{equation}
with $j_1 \leq j_2 \leq \ldots \leq j_m$ indicating the
single--particle levels occupied by the $m$ Bosons. Another equivalent
definition of the states $| \mu \rangle$ is obtained by choosing
partitions of $m$, i.e., sequences of positive integers $\{n_1, n_2,
\ldots, n_r\}$ with $r\leq l$ and $\sum_{\alpha=1}^r n_\alpha = m$,
and sets $j'_1 < j'_2 < \ldots < j'_r$ of single--particle levels. For
every such choice, each of the levels $j'_s$, $s = 1,\ldots,r$ is
occupied by $n_s$ Bosons, and the corresponding $m$--particle state
has the form
\begin{equation}
\label{eq2.4}
| \mu \rangle = { (b_{j'_1}^{\dagger})^{n_1} \ldots
  (b_{j'_r}^{\dagger})^{n_r} \over \sqrt{ n_1 !\, n_2 !\, \ldots n_r
  !\,}} \, | 0 \rangle \ .
\end{equation}
The states $| \mu \rangle$ are normalized. States belonging to
different partitions or, for the same partition, to different
sequences $j'_1 < j'_2 < \ldots < j'_r$ of single--particle levels,
are mutually orthogonal. Comparison of Eqs.~(\ref{eq2.3})
and~(\ref{eq2.4}) yields an implicit definition of the coefficients
${\cal N} (j_1,\ldots,j_m)$. These same coefficients appear also in
Eq.~(\ref{eq2.1}). The dimension of Hilbert space is
\begin{equation}
\label{eq2.5}
N = { l + m - 1 \choose m }\ .
\end{equation}

Eq.~(\ref{eq2.4}) can be used to define classes of states. All states
in a given class have in common the same set $\{n_1, \ldots, n_r\}$ of
positive integers appearing in Eq.~(\ref{eq2.4}), irrespective of the
choice of single--particle states $j'_1, \ldots, j'_r$. By definition,
permutations of the $l$ single--particle levels leave the class
invariant. We note that for Fermions there is only one class of
states. The number $N_{\rm cl}$ of classes is given by the number of
partitions of $m=\sum n_\alpha$ subject to the constraint $n_1 \geq
n_2 \geq \ldots \geq n_r$ ($r \leq l$). The number of non--zero matrix
elements coupling the state $| \mu \rangle$ to other states depends on
the class to which $| \mu \rangle$ belongs.  Therefore, a graphical
representation using vertices for the Hilbert space vectors and links
for the non--zero non--diagonal matrix elements (cf.
Ref.~\cite{ben01b}) leads to a non--regular graph in general. However,
the subgraph defined only among states of the same class is a regular
graph, i.e., the number of links emanating from each vertex is the
same for all vertices in the subgraph.

\section{The Second Moment of the Interaction}
\label{bas}

As in the Fermionic case, the second moment of the matrix $\langle \nu
| V_k(\beta) | \mu \rangle$ is of central importance in our approach.
We show that the duality relation found for Fermions also holds for
the Bosonic ensembles. 

\subsection{The Second Moment}
\label{sec}

By virtue of the randomness of $V_k(\beta)$, the elements of the
matrix $\langle \nu | V_k(\beta) | \mu \rangle$ are random variables
with a Gaussian probability distribution and zero mean value. The
distribution is completely specified in terms of the second moment
$B^{(k)}_{\mu \nu, \rho \sigma}(\beta)$ defined by
\begin{equation}
\label{eq3.1}
B^{(k)}_{\mu \nu, \rho \sigma}(\beta) =
  \overline{\langle \mu | V_k(\beta) | \sigma \rangle \langle \rho
  | V_k(\beta) | \nu \rangle} \ .
\end{equation}
To simplify the notation, we define the operators
\begin{equation}
\label{eq3.2}
{\cal B}_{\alpha (r)}^\dagger = {b_{j_1}^\dagger \ldots b_{j_r}^\dagger
  \over {\cal N}(j_1,\ldots,j_r)} \ ,
\end{equation}
and likewise for the adjoint ${\cal B}_{\alpha(r)}$. The index
$\alpha(r)$ is a short--hand notation for the rank $r$ and for the
sequence of indices $\{j_1 \ldots j_r\}$. The matrix elements $\langle
\mu | {\cal B}_{\alpha(r)}^\dagger {\cal B}_{\alpha(r)} | \nu \rangle$
are real. Using this fact, Eqs.~(\ref{eq2.1}) and~(\ref{eq2.2}), and
the Hermitecity of $V_k(\beta)$, we obtain
\begin{eqnarray}
\label{eq3.3}
B^{(k)}_{\mu \nu, \rho \sigma} (\beta)&=&\sum_{\alpha (k), \gamma(k)}
  \langle \mu | {\cal B}_{\alpha (k)}^{\dagger} {\cal B}_{\gamma(k)} |
  \sigma \rangle \nonumber \\
& & \qquad \times \Bigl[ \langle \rho | {\cal B}_{\gamma(k)}^{\dagger}
  {\cal B}_{\alpha (k)} | \nu \rangle + \delta_{\beta 1} \langle \nu |
  {\cal B}_{\gamma(k)}^{\dagger} {\cal B}_{\alpha (k)} | \rho \rangle
  \Bigr] \ .
\end{eqnarray}
The second moment, Eq.~(\ref{eq3.3}), has the same central importance
for the theory developed here as its counterpart, Eq.~(14) of
Ref.~\cite{ben01b}, for the Fermionic ensemble. For $k = m$ and $\beta
= 2$, Eq.~(\ref{eq3.3}) reduces to $\delta_{\mu \nu} \delta_{\rho
  \sigma}$, and correspondingly for $\beta = 1$. This shows that for
$k = m$, our ensembles reduce to the GUE or GOE, respectively, with
the radius of the semicircle growing as $\sqrt{N}$.

\subsection{Duality}
\label{dua}

The duality relation establishes a connection between the second
moment of the $k$--body interaction and that of the $(m-k)$--body
interaction. The proof presented here is somewhat more general than
the one given in Ref.~\cite{ben01b}.

We consider Eq.~(\ref{eq3.3}) for the unitary case $\beta=2$. We
observe that the bra--state $\langle \mu | {\cal B}_{\alpha
(k)}^{\dagger}$ contains $(m-k)$ Bosons and yields the vacuum state
upon the action of a uniquely defined operator ${\cal B}_{\beta
(m-k)}^{\dagger}$. For the matrix element in Eq.~(\ref{eq3.3}) not to
vanish, the ket--state ${\cal B}_{\gamma (k)} | \sigma \rangle$ must
contain the same $(m-k)$ Bosons as $\langle \mu | {\cal B}_{\alpha
(k)}^{\dagger}$. In other words, this state yields the vacuum state by
the action of the operator ${\cal B}_{\beta (m-k)}$ with the ${\it
same}$ index $\beta(m-k)$. The $(m-k)$--Boson states ${\cal B}_{\beta
(m-k)}^\dagger | 0 \rangle $ are orthonormal and complete, i.e., we
have for fixed $(m-k)$ that $\sum_{\beta (m-k)} {\cal B}_{\beta
(m-k)}^{\dagger} | 0 \rangle \langle 0 | {\cal B}_{\beta (m-k)} =
1_{(m-k)}$, with $1_{(m-k)}$ the unit operator for states with $(m-k)$
Bosons. Likewise, the $k$--Boson states ${\cal B}_{\alpha (k)}^\dagger
| 0 \rangle$ are orthonormal and complete for the states with $k$
Bosons. Hence,
\begin{eqnarray}
\label{eq3.4}
B^{(k)}_{\mu \nu, \rho \sigma} (2)
& = & \sum_{\alpha (k), \gamma (k) \atop \beta (m-k), \delta (m-k)}
  \langle \mu | {\cal B}_{\alpha (k)}^{\dagger}
  {\cal B}_{\beta (m-k)}^{\dagger} | 0 \rangle \langle 0 |
  {\cal B}_{\beta (m-k)} {\cal B}_{\gamma (k)}| \sigma \rangle
  \nonumber \\
& & \qquad\qquad
  \times \langle \rho | {\cal B}_{\gamma (k)}^{\dagger}
  {\cal B}_{\delta (m-k)}^{\dagger} | 0 \rangle \langle 0 |
  {\cal B}_{\delta (m-k)} {\cal B}_{\alpha (k)}| \nu \rangle
  \nonumber \\
& & \nonumber\\
& = & \sum_{\alpha (k), \gamma (k) \atop \beta (m-k), \delta (m-k)}
  \langle \mu | {\cal B}_{\beta (m-k)}^{\dagger}
  {\cal B}_{\alpha (k)}^{\dagger} | 0 \rangle \langle 0 |
  {\cal B}_{\alpha (k)} {\cal B}_{\delta (m-k)}| \nu \rangle
  \nonumber \\
& & \qquad\qquad
  \times \langle \rho | {\cal B}_{\delta (m-k)}^{\dagger}
  {\cal B}_{\gamma (k)}^{\dagger} | 0 \rangle \langle 0 |
  {\cal B}_{\gamma (k)} {\cal B}_{\beta (m-k)}| \sigma \rangle
  \nonumber \\
& & \nonumber\\
& = & \sum_{\beta (m-k) \atop \delta (m-k)}
  \langle \mu | {\cal B}_{\beta (m-k)}^{\dagger}
  {\cal B}_{\delta (m-k)} | \nu \rangle \langle \rho |
  {\cal B}_{\delta (m-k)}^{\dagger} {\cal B}_{\beta (m-k)} |
  \sigma \rangle \ .
\end{eqnarray}
The first equality in Eq.~(\ref{eq3.4}) follows from the definition of
the second moment and the completeness of the $(m-k)$--Boson states. In
the second one, we first rearrange the order of the factors
conveniently and then use the commutation properties of the Bosonic
operators. The last line follows from the completeness of the
$k$--Boson states. Using the definition of the second moment,
Eq.~(\ref{eq3.3}), we obtain
\begin{equation}
\label{eq3.5}
B^{(k)}_{\mu\nu,\rho\sigma} (2) = B^{(m-k)}_{\mu\sigma,\rho\nu} (2) \ .
\end{equation}
This is the duality relation. The duality relation is different from
and has nothing to do with the particle--hole symmetry that is used
for treating Fermions in more than half--filled shells.

\section{Eigenvector Expansion of the Second \\ Moment}
\label{eig}

In the Fermionic case, results for the shape of the spectrum and
spectral fluctuations could be derived with the help of the
eigenvector decomposition of the second moment. A similar
decomposition exists in the case of Bosons and is derived below. A
central role in the derivation is played by the identity
\begin{equation}
\label{eq4.1}
\sum_{j_1 \leq \ldots \leq j_k} {b_{j_1} \ldots b_{j_k}
  b_{j_k}^\dagger \ldots b_{j_1}^\dagger \over \bigl[ {\cal N}(j_1,
  \ldots,j_k)\bigr]^2} = {1 \over k! \, } \sum_{j_1, \ldots , j_k}
  b_{j_1} \ldots b_{j_k} b_{j_k}^\dagger \ldots b_{j_1}^\dagger \ .
\end{equation}
This identity relates a sum over the indices $j_1, \ldots, j_k$
restricted by the condition $j_1\leq \ldots \leq j_k$ containing the
appropriate weights to the unrestricted sum over the same indices. The
weight factor $k! / \bigl[ {\cal N}(j_1,\ldots,j_k)\bigr]^2$ is
related to the number of permutations with repetitions of $k$ indices.
A proof of this identity is given in the Appendix.

We consider first the unitary ensemble. Proceeding as in the case of
Fermions, we solve the eigenvalue equation $\sum_{\rho\sigma}
B^{(k)}_{\mu \nu, \rho \sigma} C^{(sa)}_{\sigma \rho} =
\Lambda^{(s)}_{\rm B}(k) C^{(sa)}_{\mu \nu}$ and show that the
eigenvectors have the form
\begin{equation}
\label{eq4.2}
C^{(sa)}_{\sigma \rho} = \langle \sigma | {\cal B}_{\beta(s)}^\dagger
  {\cal B}_{\delta(s)} | \rho \rangle \ .
\end{equation}
Here, $s$ is the rank of ${\cal B}_{\beta(s)}^\dagger$ and ${\cal
B}_{\delta(s)}$, and $a$ labels the pair $\bigl(\beta(s), \delta(s)
\bigr)$. We assume first that no index in $\beta(s)$ equals any of the
indices in $\delta(s)$. Using Eqs.~(\ref{eq3.3}) and~(\ref{eq4.2}) in
the eigenvalue equation, we obtain
\begin{equation}
\label{eq4.3}
\sum_{\rho \sigma} B^{(k)}_{\mu \nu, \rho \sigma}(2) C^{(sa)}_{\sigma
  \rho} = \sum_{\alpha(k) \gamma(k)} \langle \mu | {\cal
  B}_{\alpha(k)}^\dagger {\cal B}_{\delta(s)} {\cal B}_{\gamma(k)}
  {\cal B}_{\gamma(k)}^\dagger {\cal B}_{\beta(s)}^\dagger {\cal
  B}_{\alpha(k)} | \nu \rangle \ .
\end{equation}
We have used that there are no common indices in $\beta(s)$ and
$\delta(s)$, so ${\cal B}_{\beta(s)}^\dagger$ and ${\cal
  B}_{\delta(s)}$ commute. To perform the summation over $\gamma(k)$
we use the identity~(\ref{eq4.1}) and the fact that $\sum_j b_j
b_j^\dagger = l + \sum_j b_j^\dagger b_j$. This yields $l + m + s - 1
\choose k$. For the sum over $\alpha(k)$ we proceed similarly, 
rearranging first the order of the operators to bring together the
operators involving $\alpha(k)$. We use an identity similar
to~(\ref{eq4.1}) to compute the summation. We obtain the factor $m-s
\choose k$ multiplying $C^{(sa)}_{\mu \nu}$. This shows that the
matrix $C^{(sa)}_{\sigma \rho}$ defined in Eq.~(\ref{eq4.2}) is indeed
an eigenvector with eigenvalue 
\begin{equation}
\label{eq4.4}
\Lambda^{(s)}_{\rm B}(k) = {m-s \choose k} {l+m+s-1 \choose k} \ .
\end{equation}
In Eq.~(\ref{eq4.4}) and below, the label B stands for Bosons; we
shall use the label F for Fermions. The eigenvalues
$\Lambda^{(s)}_{\rm B}(k)$ decrease monotonically with increasing $s$
and terminate at $s = m-k$, as for the case of Fermions. Again, this
is consistent with the GUE result for $k=m$, where only the term $s =
0$ occurs and where $\Lambda^{(0)}_{\rm B}(m) = N$.

Eigenvectors of the form~(\ref{eq4.2}) with no common index in
$\beta(s)$ and $\delta(s)$ are orthogonal. In the case where there is
a common index, a construction identical to the Fermionic case is
required. The eigenvectors can be ``lifted'' from a lower rank $s'$ to
a higher one with $s> s'$ by insertion of powers of the number
operator.  The construction of linearly independent eigenvectors
$C^{(sa)}$ proceeds as for Fermions (see Ref.~\cite{ben01b} for
details), and one finds that these eigenvectors posses the same
eigenvalues~(\ref{eq4.4}).  As in the Fermionic case, the eigenvectors
do not depend upon the rank $k$ of the interaction, only the
eigenvalues do. The construction of the eigenvectors is identical in
both cases. The dimension $D^{(s)}_{\rm B}$ of the subspace spanned by
degenerate eigenvectors characterized by $s$ is given by $D^{(0)}_{\rm
  B}= 1$ and for $s\geq 1$ by
\begin{equation}
\label{eq4.6}
D^{(s)}_{\rm B} = {l + s - 1 \choose s}^2 - {l +s-2\choose s-1}^2 \ .
\end{equation}
It follows that $\sum_{s=0}^m D^{(s)}_{\rm B} = N^2$. Therefore, the
eigenvectors form a complete basis. We impose Hermitecity on the
eigenvectors and normalize them according to
\begin{equation}
\label{eq4.7}
\sum_{\mu \nu} C^{(sa)}_{\mu \nu} C^{(tb)}_{\nu \mu} = N
  \delta_{s t}\delta_{a b}\ .
\end{equation}

\begin{figure}
\begin{minipage}{15cm}
{\psfig{figure=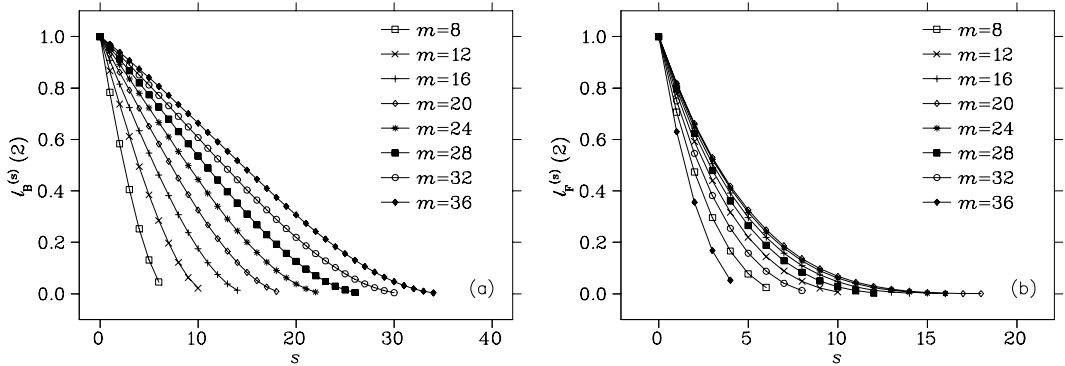,width=4.8cm,angle=90}}
\end{minipage}
\caption{The ratio $\ell^{(s)} (k) = \Lambda^{(s)}(k) /
  \Lambda^{(0)}(k)$ as a function of $s$ for (a)~Bosons and
(b)~Fermions, for $k=2$, $l=40$, and for several values of $m$. Only
non--zero eigenvalues are shown. We note the difference in horizontal
scales in the two plots.}
\label{fig1}
\end{figure}

It is instructive to compare the eigenvalues for Fermions and for
Bosons. We recall that for Fermions, the eigenvalues have a form
similar to Eq.~(\ref{eq4.4}), with the second factor on the
right--hand side replaced by $l-m+k-s \choose k$. Since the dimensions
of the Hilbert spaces for Bosons and for Fermions differ, and since
the eigenvalues depend on the dimension, a fair comparison is obtained
by normalizing the eigenvalues to the value $\Lambda^{(0)}(k)$. For
Bosons and Fermions, the ratios $\ell^{(s)}_i(k) = \Lambda_i^{(s)} (k)
/ \Lambda_i^{(0)}(k)$ with $i =$ F, B are given by
\begin{eqnarray}
\label{eq4.5a}
\ell^{(s)}_{\rm B}(k) &=& \prod_{i=1}^k \Bigl[{m+1-i-s \over m+1-i} \Bigl]
\Bigl[{l+m-i+s \over l+m-i} \Bigl] \ , \\
\label{eq4.5b}
\ell^{(s)}_{\rm F}(k) &=& \prod_{i=1}^k \Bigl[{m+1-i-s \over m+1-i} \Bigl]
\Bigl[{l-m+i-s \over l-m+i} \Bigl] \ .
\end{eqnarray}
It is easy to see from Eqs.~(\ref{eq4.5a}) and~(\ref{eq4.5b}) that
$\ell^{(s)}_{\rm F}(k) < \ell^{(s)}_{\rm B}(k)$ while $\ell^{(0)}_{\rm
  F}(k) = \ell^{(0)}_{\rm B}(k) = 1$ by definition. In the dilute limit
$k\ll m\ll l$ where the distinction between Fermions and Bosons is
irrelevant, one finds the expected result $\ell^{(s)}_{\rm F}(k)
\simeq \ell^{(s)}_{\rm B}(k)$. On the other hand, for fixed $k$ and
$l$, the difference between the Fermionic and the Bosonic case
increases with increasing particle number. For Bosons, the eigenvalues
with larger $s$ become more important. This fact suggests stronger
deviations from Wigner--Dyson statistics for Bosons than for Fermions.
This is particularly obvious for values of $m$ beyond half--filling,
$2m > l$, and $m-k>l-m$, where $\ell^{(s)}_{\rm F}(k) = 0$ for $s>l-m$
while $\ell^{(s)}_{\rm B}(k)$ is different from zero. This is
demonstrated in Fig.~\ref{fig1} which shows $\ell^{(s)}(k)$ for Bosons
and for Fermions as a function of $s$, for $k = 2$, $l = 40$ and for
several values of $m$.

For the orthogonal case we observe that the matrix $B^{(k)}(1)$ in
Eq.~(\ref{eq3.3}) consists of two terms, each one of the form of
$B^{(k)}(2)$. Combining the expressions for $\beta = 1$ and for $\beta
= 2$, we find for the second moment the eigenvector expansion
\begin{equation}
\label{eq4.8}
B^{(k)}_{\mu \nu, \rho \sigma} (\beta) = {1 \over N} \sum_{sa}
  \Lambda^{(s)}_{\rm B}(k)\Bigl[C^{(sa)}_{\mu \nu}
  C^{(sa)}_{\rho\sigma} + \delta_{\beta 1} C^{(sa)}_{\mu \rho}
  C^{(sa)}_{\nu \sigma} \Bigr] \ .
\end{equation}
This expression shows a formal analogy to the Fermionic case. However,
the dimension of Hilbert space, the values of the $\Lambda^{(s)}_{\rm
  B}(k)$, and the definitions of the eigenvectors, are all different.

\section{Low Moments of $V_k$}
\label{low}

We apply the results obtained in Sections~\ref{bas} and~\ref{eig} to
calculate the dependence on $k$, $m$, and $l$ of measures of the shape
of the average spectrum in the limit $N \to \infty$. Only in this
limit can we expect to obtain generic results. In contradistinction to
the Fermionic case, where the limit $N \to \infty$ implies $l \to
\infty$, the limit $N \to \infty$ can for Bosons also be obtained by
letting $m \to \infty$. In particular, we will be interested in this
case for fixed values of $l$ and $k$. We refer to this case as to the
dense limit. Having chosen the Bosonic ensembles in such a way that
they agree with GUE or GOE for $k = m$, and observing that for $k \ll
m \ll l$ Bosonic and Fermionic ensembles must agree, we do not expect
a qualitatively different behavior for Bosonic and Fermionic systems
as long as $m < l$. Thus, the dense limit is the really interesting
and novel case for Bosons. It has no analogue for Fermions. 

We consider three ratios that yield information about the shape of the
average spectrum. The ratio $S$ measures the fluctuations of the
center of the spectrum in units of the average width of the spectrum.
The ratio $R$ measures the relative fluctuation of the width of the
spectrum. These two quantities yield information about finite--size
effects in numerical calculations~\cite{flo01}. The third ratio is the
kurtosis $\kappa$, written as $\kappa = Q + 2$. The quantity $Q$ marks
the difference between the semicircular ($Q = 0$) and the Gaussian
shape ($Q = 1$) of the average spectrum. The definitions of $S$, $R$,
and $Q$ are given in Ref.~\cite{ben01b}. Since the eigenvectors and
the duality relation are formally similar to the Fermionic case, the
formal expressions for the three quantities $S$, $R$, and $Q$ are
identical, too. Differences are due to the changed form of the
eigenvalues, to different ranges of the parameters $k$, $m$, and $l$,
and to the following fact.

The creation and annihilation operators for Bosons are not nilpotent
like those for Fermions. Therefore, contributions of the form
$\sum_{\mu\nu} C^{(sa)}_{\mu\nu} C^{(sa)}_{\mu\nu}$ that for Fermions
are of relative order ${\cal O} (1/l^2)$ and vanish in the large $N$
limit, may become important for the Bosonic ensembles. Such terms can
be worked out explicitly exploiting the symmetry or antisymmetry of
the Hermitean eigenvectors $C_{\mu \nu}^{(sa)}$ of the second moment.
We define the ``parity'' $\pi_{a(s)}$ as zero (one) when $C_{\mu
\nu}^{(sa)}$ is symmetric (antisymmetric) in the indices $\mu, \nu$,
and have $C^{(sa)}_{\mu\nu} = (-1)^{\pi_{a(s)}} C^{(sa)}_{\nu\mu}$.
Using duality and the orthogonality relations, Eqs.~(\ref{eq3.5})
and~(\ref{eq4.7}), in $\sum_{sa} \Lambda^{(s)}(m) \sum_{\mu\nu}
C^{(sa)}_{\mu\nu} C^{(sa)}_{\mu\nu}$, we obtain the sum rule
$\sum_{sa} (-1)^{\pi_{a(s)}} = N$. From this sum rule, it is easy to
derive the following important identity for the partial sum
$d^{(s)}_{\rm B} = \sum_{a} (-1)^{\pi_{a(s)}}$,
\begin{equation}
\label{eq5.1}
d^{(s)}_{\rm B} = {l+s-1 \choose s} -
  {l+s-2 \choose s-1} = {l+s-2 \choose s} \ .
\end{equation}
Using Eq.~(\ref{eq5.1}) and the result for the Fermionic ensembles,
we obtain
\begin{eqnarray}
\label{eq5.2}
S(k,m,l) &=& {(1+\delta_{\beta 1})} \Lambda^{(0)}_{\rm B}(m-k)
  \Bigl[{\rm tr} \overline{[V_k(\beta)]^2} \Bigr]^{-1} \ , \\
  \nonumber \\
\label{eq5.3}
R(k,m,l) &=& 2 \Biggl\{ (1 + \delta_{\beta 1}) \sum_{s} \bigl(
  \Lambda^{(s)}_{\rm B} \bigr)^2 D^{(s)}_{\rm B} + 2 \delta_{\beta 1}
  \sum_{s=0}^k \bigl(\Lambda^{(s)}_{\rm B} (m-k) \bigr)^2 
  d^{(s)}_{\rm B} \Biggr\} \nonumber \\
  & & \qquad \times \ \Bigl[{\rm tr} \overline{[V_k(\beta)]^2} 
  \Bigr]^{-2} \ ,
\end{eqnarray}
and
\begin{eqnarray}
\label{eq5.4}
Q(k,m,l) &=& {1\over N} \Biggl\{ \sum_{s=0}^{{\rm min}(m-k,k)}
  \Lambda^{(s)}_{\rm B}(k) \Lambda^{(s)}_{\rm B}(m-k) \bigl[ 
  D^{(s)}_{\rm B} + 2 \delta_{\beta 1} d^{(s)}_{\rm B} \bigr]
  \nonumber \\
& + & \delta_{\beta 1} \Bigl[ \sum_{s=0}^{m-k} \bigl( 
  \Lambda^{(s)}_{\rm B}(k) \bigr)^2 d^{(s)}_{\rm B} - {2 \over N} 
  \Bigl( \sum_{s} \Lambda^{(s)}_{\rm B} d^{(s)}_{\rm B} \Bigr)^2 
  \nonumber \\
& + & {2 \over N^2} \sum_{(sa),(tb)} (-)^{\pi_{a(s)}+\pi_{b(t)}}
  \Lambda^{(s)}_{\rm B} \Lambda^{(t)}_{\rm B} {\rm tr} 
  (C^{(sa)} C^{(sa)} C^{(tb)} C^{(tb)}) \Bigr] \Biggr\} \nonumber \\
& & \ \times \ \Bigl[{1 \over N} {\rm tr} \overline{[V_k(\beta)]^2} 
  \Bigr]^{-2} \ .
\end{eqnarray}
The width of the spectrum is given by
\begin{equation}
\label{eq5.5}
{1 \over N} {\rm tr} \overline{[ V_k(\beta)]^2} =
  \Lambda^{(0)}_{\rm B} (k) + {\delta_{\beta 1} \over N}
  \sum_{s} \Lambda^{(s)}_{\rm B} d^{(s)}_{\rm B} \ .
\end{equation}
In Eqs.~(\ref{eq5.2})--(\ref{eq5.5}), traces are always taken in the
Hilbert space of the many--body states $|\mu\rangle$. We have
simplified the notation by omitting the dependence on $k$ or $m-k$ of
the eigenvalues $\Lambda^{(s)}_{\rm B}$ in the terms which are
invariant under the exchange $k \leftrightarrow m-k$ because of the
duality relation.

We consider the large $N$ limit, first realized by letting $l \to
\infty$, with $m/l$ fixed. The results are similar to the Fermionic
case: For $k\ge 1$ both $S \to 0$ and $R \to 0$. The convergence to
the limit is particularly slow when both $k$ and the filling factor
$m/l$ are fixed. In this case, $S \propto l^{-k} \propto (\ln N)^{-k}$
and $R \propto l^{-2k} \propto (\ln N)^{-2k}$ while for the canonical
ensembles ($k=m$) the limit is approached as $S(m,m,l) \propto 1/N^2$ 
and $R(m,m,l) \propto 1/N^2$. The behavior of $Q$ suggests that the
shape of the spectrum coincides with the semicircle for $2k > m$.
The semicircle
undergoes a smooth transition to Gaussian shape for $2k \le m$. The
critical value of $k$ where this transition takes place is associated,
as in the case for Fermions, to duality.

We turn to the dense limit, letting $m \to \infty$ with both $k$ and
$l$ fixed. We find
\begin{eqnarray}
\label{eq5.6}
\lim_{m\to \infty}S(k,m,l)&=&{(1+\delta_{\beta 1}){2k \choose k}
  {l+k-1 \choose k}^{-1} \over {2k \choose k} + \delta_{\beta 1} 
  \sum_{s=0}^k {2k \choose k+s} {l+k+s-1 \choose k+s}^{-1} 
  d^{(s)}_{\rm B}}\ , \\ 
  \nonumber \\  \nonumber \\
\label{eq5.7}
\lim_{m\to \infty} R(k,m,l) &=& { 2\sum_{s=0}^k [{2k \choose k+s}
  {l+k+s-1\choose k+s}^{-1}]^2 \bigl[D^{(s)}_{\rm B} +\delta_{\beta 1}
  (D^{(s)}_{\rm B} + 2d^{(s)}_{\rm B}) \bigr] \over \Bigl[{2k \choose
  k} +\delta_{\beta 1} \sum_{s=0}^k {2k \choose k+s} {l+k+s-1 \choose
  k+s}^{-1} d^{(s)}_{\rm B}\Bigr]^2} \ , \nonumber \\ \\ \nonumber \\
\label{eq5.8}
\lim_{m\to \infty} Q(k,m,l) &=& \sum_{s=0}^k {2k \choose k}^{-1} 
  {2k \choose k+s} {l+k+s-1 \choose k+s}^{-1} D^{(s)}_{\rm B} 
  \nonumber\\
  &=& 1 \qquad (\beta=2).
\end{eqnarray}

The identity used to obtain the second equality in Eq.~(\ref{eq5.8})
is derived in Ref.~\cite{rupp01}. Eq.~(\ref{eq5.8}) applies only in
the unitary case ($\beta=2$) and implies that in the dense limit, the
average spectrum has Gaussian shape. We have been unable to extend
this result to the orthogonal case ($\beta=1$). This is because we did
not succeed in calculating the term ${\rm tr} (C^{(sa)} C^{(sa)}
C^{(tb)} C^{(tb)})$ appearing in Eq.~(\ref{eq5.4}). On physical
grounds, however, we expect an equation analogous to Eq.~(\ref{eq5.8})
to be valid also for the BEGOE($k$). We conclude that the shape of the
average spectrum has Gaussian form. This is in keeping with the
results of Ref.~\cite{kot80}.

A more important and surprising result lies in the fact that the
right--hand sides of Eqs.~(\ref{eq5.6}) and~(\ref{eq5.7}) do not
vanish: The fluctuations of the centroids and of the variances of
individual spectra do not vanish asymptotically. This feature differs
from the behavior both of canonical random--matrix theory and of the
embedded Fermionic ensembles. We are led to the important conclusion
that the Bosonic ensembles are not ergodic in the dense limit $m \to
\infty$ with $k$ and $l$ fixed: Unfolding the spectra by taking the
ensemble average, or the spectral average over a single member of the
ensemble, will yield different results. We attribute this non--ergodic
behavior of the ensemble to the slow rate at which the eigenvalues
$\Lambda^{(s)}_{\rm B}(k)$ decrease as $s$ increases. Physically, it
is a consequence of the fact that the number of independent random
variables $K_\beta$ in the ensemble does not grow with $m$ but stays
fixed with $k$ and $l$. We return to this point in Section~\ref{num}.

\section{Supersymmetry Approach}
\label{sup}

As in the case of Fermions, the eigenvector expansion~(\ref{eq4.8})
permits us to use the supersymmetry approach. We shall not give a
self--contained presentation of this technique in the present
context. There are review papers which describe it extensively. We
mention for instance, Refs.~\cite{efe83} and~\cite{ver85}. Moreover,
the supersymmetry technique was also described in some detail in our
preceding paper on Fermions~\cite{ben01b}. Suffice it to say that the
resulting expressions are formally the same as in the Fermionic
case. Because of specific properties of the eigenvalues $\Lambda_{\rm
B}^{(s)}(k)$, in particular in the dense limit, new contributions
appear in the lowest--order terms in the loop expansion. These
contributions may be finite and, thus, imply departures from RMT
behavior. In the dense limit, such behavior must be expected because
of the non--ergodic features found in Section~\ref{low}.

\subsection{Saddle Point}
\label{sad}

We summarize the main steps of the supersymmetric approach for the
one-- and two--point functions; details may be found in
Ref.~\cite{ben01b}. After averaging over the ensemble, the integrand
of the generating functional contains an exponential whose argument
is a sum of squares of bilinear forms in the integration variables
involving the matrix $B^{(k)}$. We use the eigenvector
expansion~(\ref{eq4.8}) and perform the Hubbard--Stratonovich
transformation. For each eigenvector $C^{(sa)}$, this introduces a
supermatrix $\sigma^{(sa)}$ of composite variables. We use the
saddle--point approximation. We define $E=(E_1 +E_2)/2$ and $\epsilon
= E_2 - E_1$, where $E_1$ and $E_2$ are the energy arguments of the
advanced and retarded Green functions, respectively. Omitting small
terms of order $1/N$, we find the coupled saddle--point equations
\begin{equation}
\label{eq6.1}
\sigma^{(s a)} = \frac{1}{N} {\rm tr}_{\mu} \biggl( [ E - \sum_{t b}
  \lambda_{\rm B}^{(t)}(k) \sigma^{(t b)} C^{(t b)} ]^{-1}
  \lambda_{\rm B}^{(s)}(k) C^{(s a)} \biggr) \ ,
\end{equation}
where $\lambda_{\rm B}^{(s)}(k)$ is the positive square root of
$\Lambda_{\rm B}^{(s)}(k)$. As in the Fermionic case, this set of
equations is solved iteratively by introducing the variable $X_{\rho
  \sigma}=\sum_{sa} \lambda_{\rm B}^{(s)}(k) \sigma^{(sa)}
C^{(sa)}_{\rho\sigma}$ and using the identity $\overline{G(E)} = [E - 
X]^{-1}$ for the averaged Green function. We find that $X$ obeys the
saddle--point equation $X=\Lambda_{\rm B}^{(0)}(k) [E-X]^{-1}$. Hence,
the solution diagonal in the superindices has the form $X^{\rm
diag}_{\mu \nu}=\delta_{\mu\nu} \lambda_{\rm B}^{(0)}(k) \tau^{(0)}$,
with
\begin{equation}
\label{eq6.2}
\tau^{(0)} = \frac{E}{2 \lambda^{(0)}_{\rm B}(k)} \pm i \sqrt{1 -
  \biggl(\frac{E}{2 \lambda^{(0)}_{\rm B}(k)}\biggr)^2} \ ,
\end{equation}
for $|E|\le 2\lambda_{\rm B}^{(0)}$. The $\pm$ signs refers to the
retarded ($G^-(E)$) and advanced ($G^+(E)$) case, respectively. This
implies that the solutions of Eq.~(\ref{eq6.1}) reduce to $\tau^{(0)}$
for $s=0$, while $\sigma^{(sa)}=0$ for $s\ge 1$. For the one--point
function this shows that, within the range of validity of the
saddle--point approximation, the semicircle describes the shape of the
average spectrum.

As in the case of the canonical ensembles, the invariance of the
effective Lagrangean under general pseudounitary transformations
implies that the two--point function possesses not a single saddle
point but a saddle--point manifold which is generated by $T^{-1}
\sigma^{(0)} T$. The matrices $T$ parameterize the coset space
UOSP$(1,1/1,1) / [$UOSP$(1,1) \otimes $UOSP$(1,1)]$, and
$\sigma^{(0)}$ is a diagonal supermatrix of dimension four with
entries given by Eq.~(\ref{eq6.2}) in the usual way~\cite{ver85}.
Using this and the saddle--point solution in the expression of the
effective Lagrangean, the first--order term in $\epsilon$ takes the
canonical form
\begin{equation}
\label{eq6.4}
- \frac{i \pi \epsilon}{d} {\rm trg} ( L T^{-1} L T ) \ .
\end{equation}
Here $d$ is the average level spacing and the diagonal supermatrix $L$
(defined in Ref.~\cite{ver85}) distinguishes the retarded and advanced
cases. The argument also carries through for higher--point correlation
functions. Eq.~(\ref{eq6.4}) shows that within the range of validity
of the saddle--point approximation, the spectral fluctuations of the
BEGUE($k$) are identical to those of the GUE.

\subsection{Corrections of the Loop Expansion}
\label{loop1}

To determine the range of validity of the saddle--point solution in
the limit $N \rightarrow \infty$, we use the loop expansion, a
power--series expansion of the effective Lagrangean around the
saddle--point, followed by an expansion of the resulting exponential.
Again, the formal equivalence between the Fermionic and Bosonic
eigenvector expansions permits a straightforward transcription of the
results for Fermions.

For the one--point function, the first correction term in the loop
expansion yields
\begin{equation}
\label{eq6.5}
|\overline{G^+(0)}| \leq {N | {\rm trg} J| \over \lambda_{\rm B}^{(0)}(k)}
  \Bigl( 1 + Q(k,m,l) \Bigr) \ .
\end{equation}
From Eq.~(\ref{eq6.5}) we conclude that for $l\to \infty$, the shape
of the spectrum is the semicircle for $2k > m$, and finite corrections
arise for $2k \le m$ (see Section~\ref{low}). These results coincide
with those for the Fermionic ensembles. In particular, in the dilute
limit a Gaussian shape of the spectrum is obtained. This holds also
true in the dense limit $m\to\infty$ with $k$ and $l$ fixed as given
by Eq.~(\ref{eq5.8}).

For the two--point function we drop terms that vanish as $1/N$ for $N
\to \infty$. The resulting expression can be written as an
exponential which contains two non--trivial terms. The first one
contains the expression
\begin{equation}
\label{eq6.6}
- \frac{i \pi \epsilon} {d_{\rm l}} {\rm trg} ( L T^{-1} L T ) \ ,
\end{equation}
where $d_{\rm l}$ is a modified level spacing which includes the terms
that lead to the bound given in Eq.~(\ref{eq6.5}) for the average Green
function. More interesting is the second term which, as in the
Fermionic case, is proportional to
\begin{equation}
\label{eq6.7}
R(k,m,l) \biggl( {1 \over d_{\rm l}} \, {\rm trg}
  \bigl[ {1\over 2} \epsilon L T L T^{-1} + L T J T^{-1} \bigr]
  \biggr)^2 \ .
\end{equation}
Kravtsov and Mirlin~\cite{kra94} have shown that terms of this form
produce deviations from universal behavior. The strength of the
fluctuations which arise from this term is proportional to the
ratio $R(k,m,l)$ given in Eq.~(\ref{eq5.3}). We recall that $R$
measures the relative fluctuations of the width of the spectrum. While
in the dilute limit this quantity tends slowly to zero as $N \to
\infty$, in the dense limit it attains a constant value. Such finite
corrections imply that the saddle--point solution is no longer
appropriate in the dense limit, and deviations from Wigner--Dyson
statistics must occur.

We conclude that in the dense limit, BEGUE($k$) possesses spectral
fluctuations which deviate significantly from the predictions of RMT.
As mentioned above, we attribute this feature to the fact that the
number of independent random variables of the ensemble remains the
same as $m$ increases, see Section~\ref{num}. Alternatively, we
attribute this result to the rather slow decrease of the eigenvalues
$\Lambda_{\rm B}^{(s)}$ with increasing $s$. Contributions other than
the canonical term $s = 0$ are relevant and give rise to important
corrections for the spectral fluctuations. These features are similar
to those characterizing the dilute limit. Here, the spectral
correlations become Poissonian. In that case, however, the function
$R(k,m,l)$ vanishes asymptotically, albeit very slowly. Therefore, the
first loop correction of the supersymmetry method is not able to
provide solid evidence for non--RMT spectral fluctuations in the
dilute limit. However, in this limit another method is available: The
extension of the binary correlation approximation of Mon and
French~\cite{mon75} developed in Ref.~\cite{ben01b}. This method works
equally for Bosons and for Fermions (in the dilute limit, the
distinction between the two cases disappears) and, thus, yields
identical results: The spectral density acquires Gaussian shape, and
the spectral fluctuations become Poissonian. 

\section{Numerical Results: the Dense Limit}
\label{num}

In the dense limit, there are no analytical techniques to predict
spectral fluctuations. Moreover, the Bosonic ensembles are not
ergodic.  Therefore, unfolding procedures based upon averaging over
the ensemble carry a question mark and spectral unfolding is the
preferred procedure to study the spectral fluctuations. In this
Section, we present numerical results on the spectral fluctuation
properties of BEGOE($k$) for $k = 2$ in the dense limit. We exhibit
the difference between spectral unfolding and unfolding by using the
average over the ensemble. For brevity, the latter method is called
ensemble unfolding.

Numerically, the dense limit is conveniently modeled by taking $l =
2$. The dimension of Hilbert space is then given by $N = m + 1$ (cf.
Eq.~(\ref{eq2.5})), and the situation $m \gg l$ can be easily attained
numerically. Furthermore, for $l = 2$ and $k = 2$ the number of
independent random variables of the ensemble is $K_1 = 6$. We
emphasize that $K_1$ is independent of the number $m$ of Bosons. The
many--particle states $| \mu_n \rangle$ are written as $(m-n, n)$
where $m-n$ and $n$ indicate the number of Bosons occupying the first
and the second single--particle state, respectively. The
many--particle states $| \mu_n \rangle$ are arranged in a sequence
$(m,0)$, $(m-1,1), \dots, (0,m)$ of increasing values of $n$. Then,
the Hamiltonian matrix attains a band structure, with non--zero matrix
elements on the main diagonal and on the $k$ closest diagonals. We
restrict ourselves to the physically most relevant case $k = 2$.

\subsection{Ensemble Unfolding versus Spectral Unfolding}
\label{unfold}

The evaluation of measures of spectral fluctuations requires that
individual spectra be unfolded: The average level spacing must be
constant. Spectra can be unfolded in two ways. Spectral unfolding
consists in handling each spectrum independently. One looks for a
transformation which yields the value unity for the average level
spacing. This transformation need not be the same for different
spectra. Ensemble unfolding uses the average spectral density
(obtained by averaging all spectra in the ensemble) to find a single
transformation which is then applied to all spectra of the ensemble.

Spectral unfolding is typically done by fitting a polynomial to the
staircase function of each realization of the ensemble. In the results
presented below, we have determined the degree of the polynomial for
each realization by taking the value which gives the first minimum of
the associated $\chi^2$ when increasing the degree starting from 1.
The maximum degree considered was 20. Typically the first minimum was
found around 11.

Ensemble unfolding is usually preceded by removing the fluctuations of
the center and of the width~\cite{man84,flo01}. All spectra are
recentered and rescaled so as to have a common center and width. This
procedure is essentially equivalent to the more sophisticated method
used in Refs.~\cite{bro81,patel00,lab90} which corrects higher moments
of the distribution and relies strongly on the Gaussian shape of the
average spectrum. The recentered and rescaled individual spectra are
then superposed to obtain the ensemble--averaged spectral density and
the staircase function for the ensemble. The latter is fitted with a
polynomial, and each spectrum of the ensemble is unfolded using this
polynomial.  However, for the Bosonic ensembles in the dense limit,
this procedure yields a non--Gaussian average level density as a
consequence of the non--ergodic behavior of the ensemble.  Therefore,
we did not recenter and rescale the individual spectra for
constructing the ensemble--averaged spectral density and the
corresponding staircase function. We have used a polynomial of degree
$11$ for fitting the average staircase function. The stability of the
results has been checked by varying the degree of the polynomial and
the size of the bins used to construct the average spectral density.

\begin{figure}
\begin{minipage}{15cm}
  \centerline{\psfig{figure=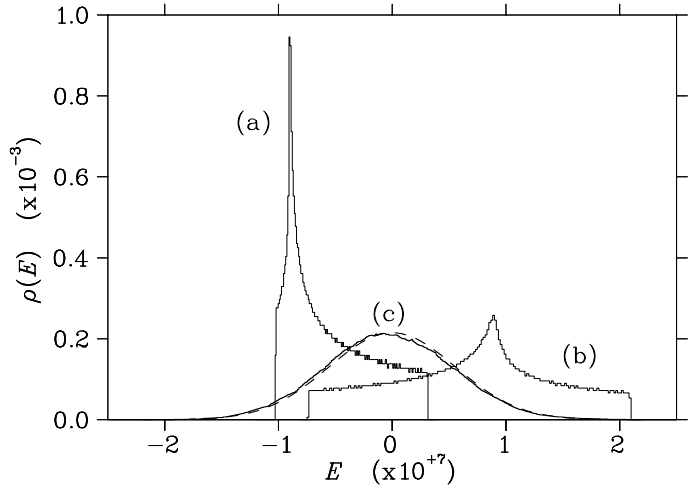,angle=90,width=9cm}}
\end{minipage}
\caption{Spectral density of the BEGOE(2) for $l = 2$ and
  $m = 3000$, normalized to the dimension $N = 3001$ of Hilbert space.
  The curves (a)~and (b)~show the level densities of two individual
  realizations of the ensemble; curve (c)~shows the ensemble--averaged
  spectral density. The dashed curve corresponds to a Gaussian shape
  of the spectrum.}
\label{fig2}
\end{figure}

Fig.~\ref{fig2} shows the spectral densities of two realizations of
the ensemble and the ensemble--averaged spectral density obtained from
$1512$ realizations of the ensemble for $m = 3000$. The integral over
each histogram is normalized to $3001$, the dimension of Hilbert
space. In the Figure, we have also plotted a Gaussian density (dashed
curve), whose width is given by the theoretical prediction
Eq.~(\ref{eq5.5}).  Comparison of the ensemble--averaged density with
this curve confirms the prediction of a Gaussian shape of the average
spectrum. The striking differences between each of the two spectral
densities and between those and the ensemble--averaged spectral
density illustrate the non--ergodic character of the Bosonic ensemble
for $m \gg l$. The Gaussian form of the average spectrum arises as a
consequence of the Central Limit Theorem. It has no bearing on
individual spectral shapes.

\subsection{Spectral Statistics}
\label{stat}

We turn to the spectral fluctuation properties of BEGOE($2$). We have
analyzed $1512$ members of the ensemble for $m = 3000$. After
fitting the polynomials, including all levels of each realization for
both spectral and ensemble unfolding, we only considered $60\%$ of all
levels, i.e., $1800$ levels for each realization. These were the
levels closest to the center of the spectrum.

\begin{figure}[t]
\begin{minipage}{15cm}
\centerline{\psfig{figure=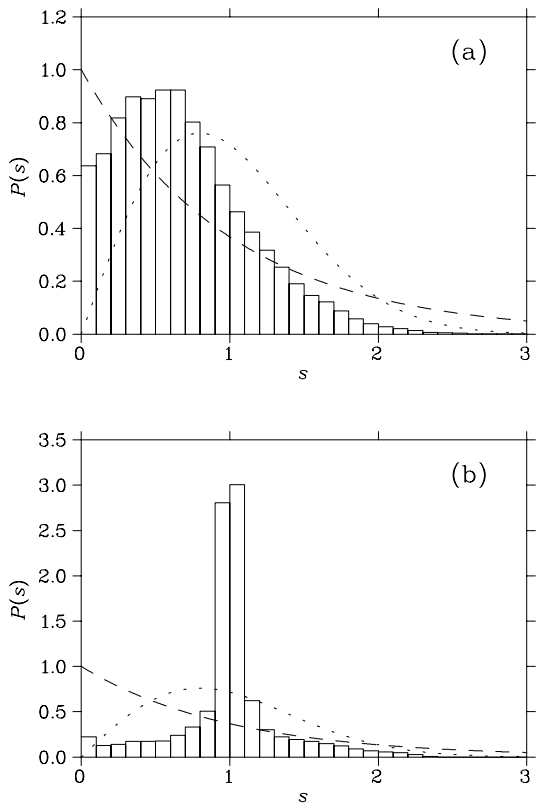,angle=90,width=7cm}}
\end{minipage}
\caption{Nearest--neighbor spacing distribution $P(s)$ obtained by
(a)~ensemble unfolding and (b)~spectral unfolding. The dotted curves
correspond to the Wigner surmise and the dashed ones to the Poisson
distribution. We note the different vertical scales used in the frames.}
\label{fig3}
\end{figure}

In Fig.~\ref{fig3} we present the nearest--neighbor spacing
distribution $P(s)$ obtained by ensemble unfolding and by spectral
unfolding. In both cases the distribution $P(s)$ corresponds neither
to the Wigner surmise nor to a Poisson distribution. In the case of
ensemble unfolding the level repulsion characteristic of the GOE is
clearly lost. On the other hand, the tail of the distribution decays
rather rapidly in comparison to the exponential decay of the Poisson
distribution. For spectral unfolding the distribution $P(s)$ is
dominated by a prominent peak centered at $s=1$. This suggests that
individual spectra have an almost constant level spacing, i.e., are
nearly of picket--fence type (one--dimensional harmonic oscillator).

\begin{figure}[t]
\begin{minipage}{15cm}
\centerline{\psfig{figure=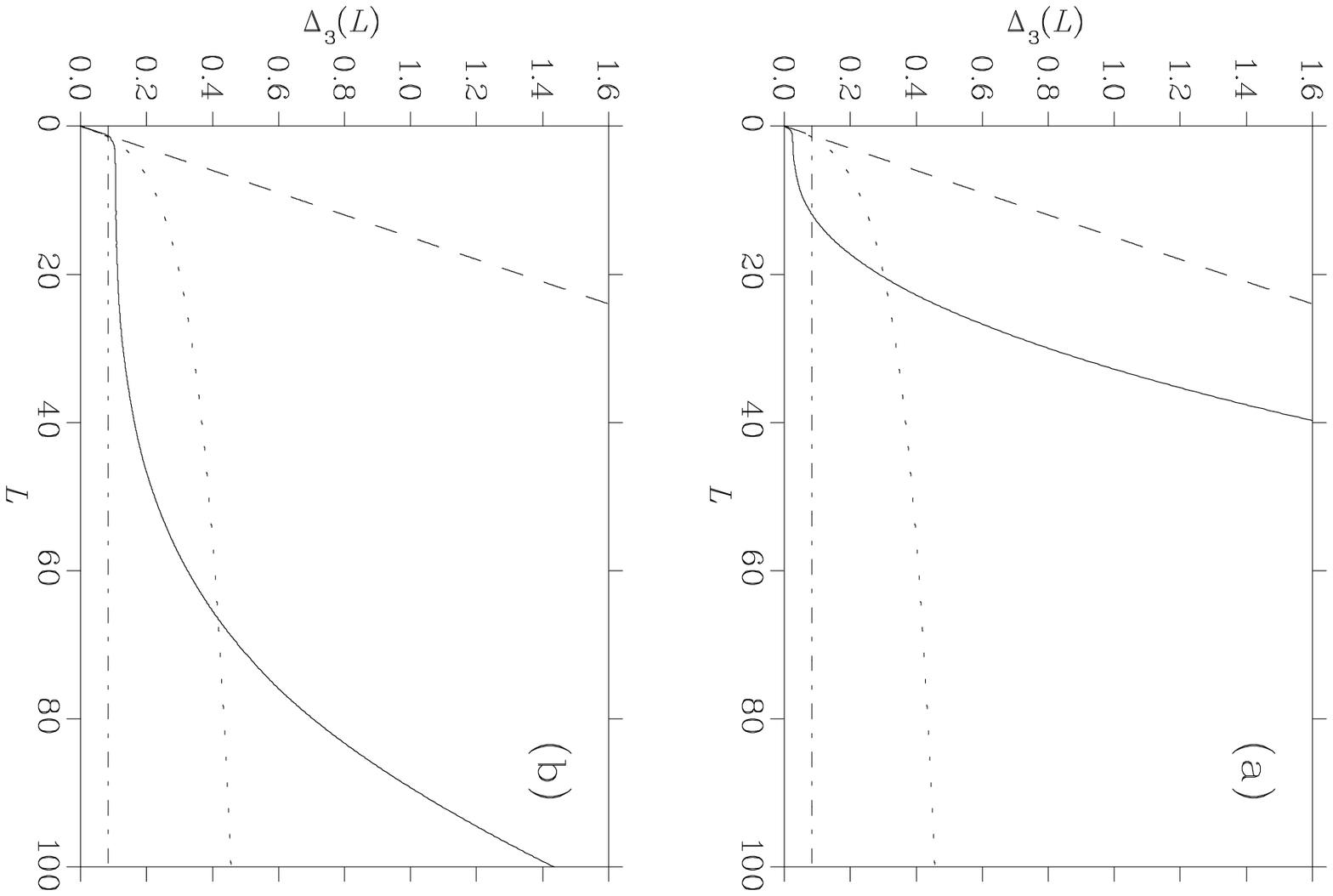,angle=90,width=7cm}}
\end{minipage}
\caption{
The $\Delta_3$--statistics (solid lines) measured at the centers of
the spectra after (a)~ensemble unfolding and (b)~spectral unfolding. For
comparison we have plotted the results for the GOE (dotted curve), for
a Poissonian spectrum (dashed line), and for a picket--fence spectrum
(dotted--dashed line).}
\label{fig4}
\end{figure}

The two--point correlations of the spectra are characterized by the
$\Delta_3$--statistics of Dyson and Mehta measured at the centers of
the spectra and shown in Fig.~\ref{fig4}. We have computed
$\Delta_3(L)$ directly from the spectra using the expression derived
by Bohigas and Giannoni~\cite{boh75}. For comparison we have also
plotted the results for a GOE spectrum $\overline{\Delta_3}(L) \sim
[\ln (2\pi L) + \gamma - \pi^2/8 -5/4]/\pi^2$, an uncorrelated
(Poisson) spectrum $\overline{\Delta_3}(L) = L/15$, and a
picket--fence spectrum $\overline{\Delta_3}(L) = 1/12$ (see
Ref.~\cite{boh75} for details). We have checked that the increase 
of $\Delta_3(L)$ with $L$ for small values of $L$ is approximately 
linear in both panels of Fig.~\ref{fig4}. Ensemble unfolding
(Fig.~\ref{fig4}(a)) yields a $\Delta_3$--statistics which deviates
rapidly from GOE behavior, and increases almost linearly. This is
consistent with the expected deviations from RMT predictions found in
Section~\ref{sup}. In the case of spectral unfolding, $\Delta_3 (L)$
is almost constant up to $L \sim 20$. This result again suggests a
tendency of individual spectra to have a picket--fence--like behavior.
Beyond this point, the $\Delta_3(L)$ grows albeit less rapidly than
after ensemble unfolding.

\subsection{Structure of Individual Spectra and Wave Functions}
\label{indiv}

The results presented above show the deviations from RMT behavior
expected after ensemble unfolding. For spectral unfolding the results
suggest that spectra of individual realizations have a very rigid
structure with an almost constant spacing in some regions of the
spectrum. We now describe in more detail the structure of individual
spectra. While we do not have a full analytical understanding for the
tendency towards picket--fence behavior, we believe that the following
observations are related to the small number of independent random
variables of the Bosonic ensemble for $m \gg l$, and to the associated
graph structure.

\begin{figure}[t]
\begin{minipage}{15cm}
\centerline{\psfig{figure=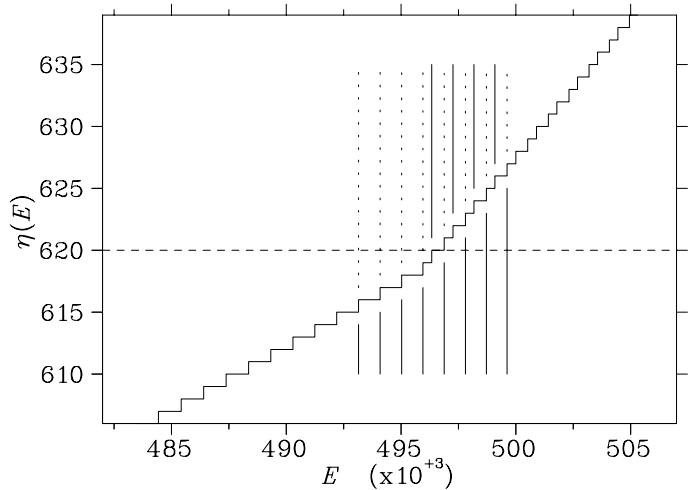,angle=90,width=9cm}}
\end{minipage}
\caption{Detail of the staircase function $\eta(E)$ for a typical
realization of BEGOE(2) with $m = 1000$ and $l = 2$. The vertical
lines help to show how two (almost) equidistant spectra overlap around
the level $620$. We note the change in the density of states after
this level.} 
\label{fig5}
\end{figure}

Fig.~\ref{fig5} shows the staircase function $\eta(E)$ for a typical
realization of the ensemble ($k = 2$, $l = 2$) for $m = 1000$. The
spectrum is dominated by levels with almost constant spacing. The
staircase function typically displays one or more points where an
abrupt change in the density of states takes place: The spectrum has
(almost) constant spacing up to the level $619$. This behavior
suddenly changes at the level $620$. While the spacing between
neighboring levels is no longer constant, the spacing between
next--to--nearest--neighbors is and is almost the same as it was
before level $620$ appeared. This is illustrated by the vertical lines
plotted above and below the staircase function. We remark that these
observations imply non--stationary properties of the spectra.

These results suggest that the spectrum of an individual realization
consists of pieces of overlapping segments of spectra with almost
constant level spacings. The kink at level $620$ in Fig.~\ref{fig5}
marks the left edge of the overlap region. Such spectra would lead to
results like those in Figs.~\ref{fig3}(b) and~\ref{fig4}(b). We notice
that if we were to use for spectral unfolding a piecewise polynomial
fit instead of a polynomial fit covering the whole spectrum, the rigid
features of the spectrum would appear in an even more convincing form
in the spectral statistics. We have not pursued this point further.

\begin{figure}
\begin{minipage}{15cm}
\centerline{\psfig{figure=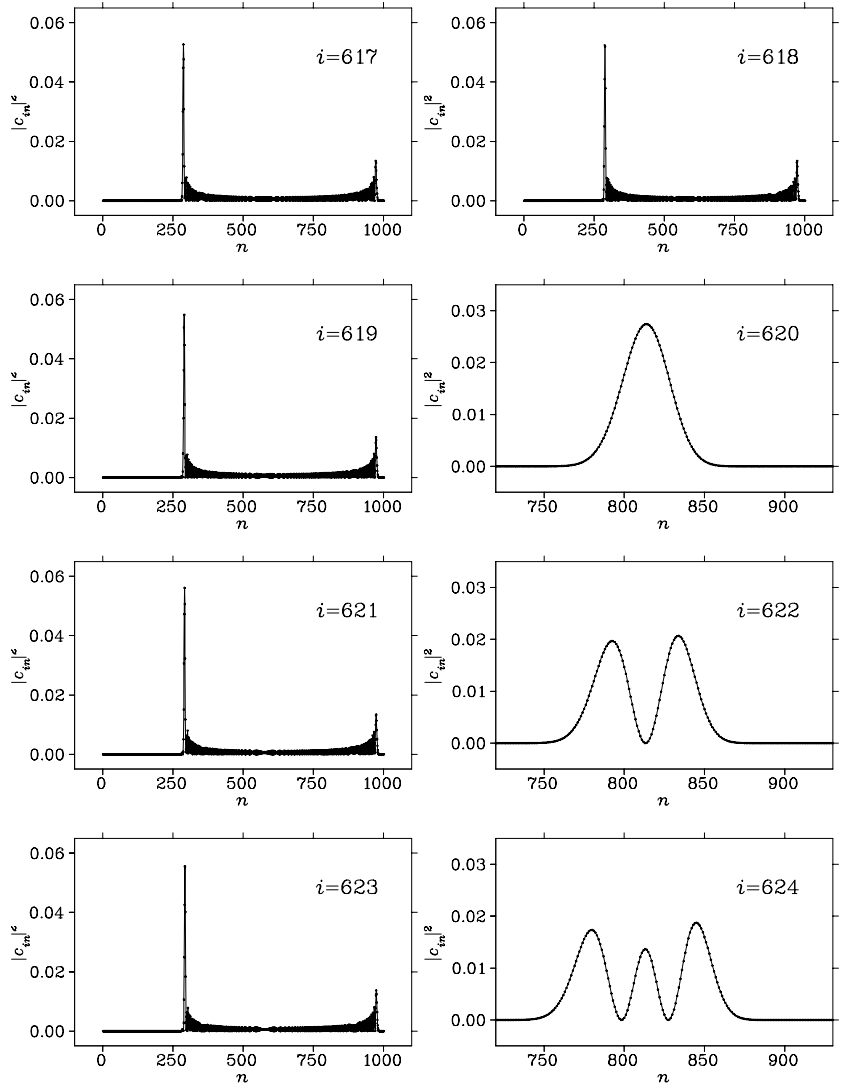,width=11cm}}
\end{minipage}
\caption{
Probabilities $|c_{in}|^2$ for eigenvectors belonging to eigenvalues
in the vicinity of the kink shown in Fig.~\ref{fig5}. The two
overlapping segments of nearly equidistant levels are easily 
distinguished by the structure of the eigenfunctions.}
\label{fig6}
\end{figure}

If the spectra do possess the structure proposed above, eigenfunctions
belonging to different segments of the spectrum should differ
qualitatively. To verify this hypothesis, we write the eigenvectors $|
i \rangle$ as linear combinations of the ordered many--body basis
states $| \mu_n \rangle$,
\begin{equation}
\label{eq8.1}
| i \rangle = \sum_{n=0}^m c_{in} | \mu_n \rangle \ .
\end{equation}
In Fig.~\ref{fig6} we plot the probabilities $|c_{in}|^2$ of
eigenvectors $| i \rangle$ belonging to eigenvalues in the vicinity of
the kink shown in Fig.~\ref{fig5}. Eigenvectors up to $i = 619$ behave
similarly and are somewhat extended, although the distribution of the
intensities clearly deviates from the Porter--Thomas distribution
expected from RMT. This behavior of the eigenfunctions changes
abruptly at $i = 620$ where the eigenvector is a rather ``coherent''
superposition of close--lying basis states. We emphasize the
difference in the horizontal scales used in Fig.~\ref{fig6} for $i =
619$ and $i = 620$, for instance. The eigenvectors with $i = 621, 623$
display the same behavior as the eigenvectors up to $i=619$, and thus
correspond to the first segment of the equidistant spectra. The
eigenvectors with $i = 620, 622, 624$ are much more localized in Fock
space. They differ in the number of intensity oscillations. As $i$
increases, so does this number, and the levels on this sequence become
more and more delocalized. At some point the spread of the
eigenfunctions in the second segment is indistinguishable from that in
the first one.

\section{Conclusions}
\label{con}

We have studied the shape of the average spectrum and the spectral
fluctuations of the Bosonic embedded ensembles BEGOE($k$) and
BEGUE($k$) in the limit of infinite matrix dimension, attained by
letting either the number $l$ of degenerate single--particle states go
to infinity, or by letting the number $m$ of Bosons go to infinity. In
the first case, the results are qualitatively the same as for the
Fermionic embedded ensembles~\cite{ben01b}: For sufficiently high rank
$k$ of the random interaction ($2k > m$), these ensembles behave
generically. The average spectrum has semicircle shape, and the
spectral fluctuations obey Wigner--Dyson statistics. A smooth
transition to a different regime takes place at or near $2k = m$. This
new regime is characterized by a Gaussian spectral shape. The spectral
fluctuations also change and are not of Wigner--Dyson type for $2k
\lesssim m$; in the dilute limit $k \ll m \ll l$ the spectral
fluctuations become Poissonian. We expect that the transition away
from Wigner--Dyson statistics is more rapid in the Bosonic than in the
Fermionic case.

In the second case ($m > l$), and in particular in the dense limit,
where $l$ and the rank of the interaction $k$ remain fixed as $m \to
\infty$, we have shown that the Bosonic ensembles are not ergodic. To
the best of our knowledge, this is the first case that a
random--matrix model has been shown to be non--ergodic in the limit of
infinite matrix dimension. Moreover, we have shown that the spectral
fluctuations deviate strongly from RMT results. This finding disagrees
with the conclusions based on numerical simulations of
Refs.~\cite{man84,patel00}. We ascribe this to the fact that the ratio
$m/l$ for the parameter settings investigated both in
Ref.~\cite{man84} ($m=11$, $l=4$) and Ref.~\cite{patel00} ($m=10$,
$l=5$) were still too small to encounter deviations from Wigner--Dyson
statistics. We have studied numerically the structure of the spectra
of individual realizations of the ensemble in the dense limit using
the case $m = 3000 $, $l = 2$ and $k = 2$ as example.  In both short--
and long--range correlations we found a tendency towards a
picket--fence type of behavior. We presented evidence pointing to a
non--stationary behavior of the spectra. The spectrum of an individual
realization seems generically to consist of segments of spectra of
picket--fence type. These segments overlap. In the overlap region, the
eigenvectors can be attributed to the different segments by their
structure which is characteristic for each segment. In particular, we
have shown that some of the eigenfunctions display strong localization
in Fock space.  We cannot offer an analytical explanation for these
properties. We believe that they are caused by the small and constant
(independent of $m$) number of independent random variables $K_\beta$
and by the specific graph structure associated with the Hamiltonian
matrix of the Bosonic ensembles. This causes the Hamiltonian matrix to
attain a particular structure with highly correlated matrix
elements. In particular, we do not know how the spectra change as $l$
increases for $m/l\gg 1$.

{\bf Acknowledgment} We are grateful to O. Bohigas and T.H. Seligman
for stimulating discussions and useful suggestions. T.A. acknowledges
support from the Japan Society for the Promotion of Science.

\section*{Appendix: Proof of the Identity Eq.~(\ref{eq4.1})}
\label{ident}

We claim that
\begin{equation}
\label{eqA.1}
{\cal S}_k = \sum_{j_1\leq j_2 \leq \ldots \leq j_k} {b_{j_1} \ldots
  b_{j_k} b_{j_k}^\dagger \ldots b_{j_1}^\dagger \over \bigl[ {\cal
  N}(j_1,\ldots,j_k) \bigr]^2 }
\end{equation}
is equal to
\begin{equation}
\label{eqA.2}
{\cal T}_k = { 1 \over k\, !} \sum_{j_1, j_2 \ldots , j_k} b_{j_1}
  \ldots b_{j_k} b_{j_k}^\dagger \ldots b_{j_1}^\dagger \ .
\end{equation}
We prove the claim by induction. The claim is trivial for $k = 1$ and
$k = 2$. We assume that the assertion holds for $k$ and prove it for
$k + 1$. On the right--hand side of Eq.~(\ref{eqA.1}), written for $k
+ 1$, we use the definition of the coefficients ${\cal N}(j_1,\ldots,
j_{k+1})$. We write
\begin{equation}
\label{eqA.3}
{\cal S}_{k+1} = \sum_{\{n_i\}} {(b_{j'_1})^{n_1} \ldots (b_{j'_r})^{n_r}
  (b_{j'_r}^\dagger)^{n_r} \ldots (b_{j'_1}^\dagger)^{n_1} \over n_1 !
  \, \ldots n_r ! \, } \ ,
\end{equation}
where the sum over the non--zero integers $n_i$ is restricted by $\sum_i
n_i=k+1$. Moreover, $j'_1< \ldots < j'_r$. We arrange the terms in the
sum in Eq.~(\ref{eqA.3}) in increasing powers of $b_{j'_r}$ (and
$b_{j'_r}^\dagger$) and apply the induction hypothesis to the
remaining sums involving at most $k$ factors $b^{\dagger}_j$. Hence,
\begin{eqnarray}
\label{eqA.4}
{\cal S}_{k+1} = && \sum_{n=1}^{k+1}\, { 1 \over (k+1-n)! \, n! }
  \nonumber \\
\times & & \sum_{j_1,\ldots , j_{k+1-n} < j_{k+2-n}}
  b_{j_1} \ldots b_{j_{k+1-n}} (b_{j_{k+2-n}})^n
  (b_{j_{k+2-n}}^\dagger)^n b_{j_{k+1-n}}^\dagger \ldots
  b_{j_1}^\dagger
  \nonumber \\
&& \nonumber \\
= & & {1 \over (k+1)!} \sum_{n=1}^{k+1} {k+1 \choose n}
  \nonumber \\
\times & & \sum_{j_1,\ldots , j_{k+1-n} < j_{k+2-n}}
  b_{j_1} \ldots b_{j_{k+1-n}} (b_{j_{k+2-n}})^n
  (b_{j_{k+2-n}}^\dagger)^n b_{j_{k+1-n}}^\dagger \ldots
  b_{j_1}^\dagger \ .
  \nonumber \\
\end{eqnarray}

Consider now the sum on the right--hand side of Eq.~(\ref{eqA.2}) for
$k+1$. For each term, we identify the largest of the indices $j_1,
\ldots j_{k+1}$. We call it $j_{\rm max}$. The $b_{j_{\rm max}}$'s and
$b_{j_{\rm max}}^\dagger$'s come in powers $n_{j_{\rm max}} = 1,
\ldots, k+1$. For a fixed value $n_{j_{\rm max}}=n$, the number of
terms in the sum is $k+1 \choose n$. Relabeling the indices in the
sum, so that $j_{\rm max} \to j_{k+2-n}$, the sum can be rewritten in
the form of Eq.~(\ref{eqA.4}).

\end{document}